\documentclass[sigconf]{acmart}

\AtBeginDocument{%
  }

\copyrightyear{2024}
\acmYear{2024}
\setcopyright{acmlicensed}
\acmConference[CIKM '24]{Proceedings of the 33rd ACM International Conference on Information and Knowledge Management}{October 21--25, 2024}{Boise, ID, USA}
\acmBooktitle{Proceedings of the 33rd ACM International Conference on Information and Knowledge Management (CIKM '24), October 21--25, 2024, Boise, ID, USA}
\acmDOI{10.1145/3627673.3679766}
\acmISBN{979-8-4007-0436-9/24/10}

\usepackage{subfigure}
\usepackage{natbib}
\usepackage{enumitem}
\usepackage{balance}

\begin{document}

\title{Structural Representation Learning and Disentanglement for Evidential Chinese Patent Approval Prediction}

\author{Jinzhi Shan}
\affiliation{%
    \institution{Beijing Institute of Technology}
    \city{Beijing}
    \country{China}
    }
\email{jz_shan@bit.edu.cn}

\author{Qi Zhang}
\affiliation{%
    \institution{Tongji University}
    \city{Shanghai}
    \country{China}
    }
\email{zhangqi_cs@tongji.edu.cn}

\author{Chongyang Shi}
\authornote{Corresponding author.}
\affiliation{%
    \institution{Beijing Institute of Technology}
    \city{Beijing}
    \country{China}
    }
\email{cy_shi@bit.edu.cn}

\author{Mengting Gui}
\affiliation{%
    \institution{Beijing Institute of Technology}
    \city{Beijing}
    \country{China}
    }
\email{mt_gui@bit.edu.cn}

\author{Shoujin Wang}
\affiliation{%
    \institution{University of Technology Sydney}
    \city{Sydney}
    \country{Australia}
    }
\email{shoujin.wang@uts.edu.au}

\author{Usman Naseem}
\affiliation{%
    \institution{University of Sydney}
    \city{Sydney}
    \country{Australia}
    }
\email{usman.naseem@sydney.edu.au}

\renewcommand{\shortauthors}{Jinzhi Shan et al.}

\begin{abstract}
Automatic Chinese patent approval prediction is an emerging and valuable task in patent analysis. However, it involves a rigorous and transparent decision-making process that includes patent comparison and examination to assess its innovation and correctness. This resultant necessity of decision evidentiality, coupled with intricate patent comprehension presents significant challenges and obstacles for the patent analysis community. Consequently, few existing studies are addressing this task.
This paper presents the pioneering effort on this task using a retrieval-based classification approach. We propose a novel framework called DiSPat, which focuses on structural representation learning and disentanglement to predict the approval of Chinese patents and offer decision-making evidence. DiSPat comprises three main components: \textit{base reference retrieval} to retrieve the Top-$k$ most similar patents as a reference base; \textit{structural patent representation} to exploit the inherent claim hierarchy in patents for learning a structural patent representation; \textit{disentangled representation learning} to learn disentangled patent representations that enable the establishment of an evidential decision-making process.
To ensure a thorough evaluation, we have meticulously constructed three datasets of Chinese patents. Extensive experiments on these datasets unequivocally demonstrate our DiSPat surpasses state-of-the-art baselines on patent approval prediction, while also exhibiting enhanced evidentiality.
\end{abstract}



\maketitle

\section{Introduction}
Patent analysis, as a valuable resource for acquiring knowledge, is widely recognized as an effective method to analyze technological innovation \citep{DBLP:journals/jsis/Grimshaw91a,MiaoCHZWZZ23}, which in turn drives economic growth \citep{mikovs2023landslide}. 
In recent years, there has been a growing emphasis on intellectual property rights, leading to a significant increase in the total number of patent applications. However, this surge has also resulted in an increased workload for patent examiners, leading to longer examination cycles and a substantial backlog of pending patent applications. This delay significantly impacts the predictability of market-oriented innovation.
Additionally, patents encompass specialized technical knowledge from various industries, making it challenging for patent examiners to master the vast knowledge reserves.
Consequently, the automation of patent examination and approval is seen as an inevitable future trend to assist in expediting and enhancing the examination process~\citep{ebrahim2018automation}.

\begin{figure}[htbp]
	\centering
        \includegraphics[width=\linewidth]{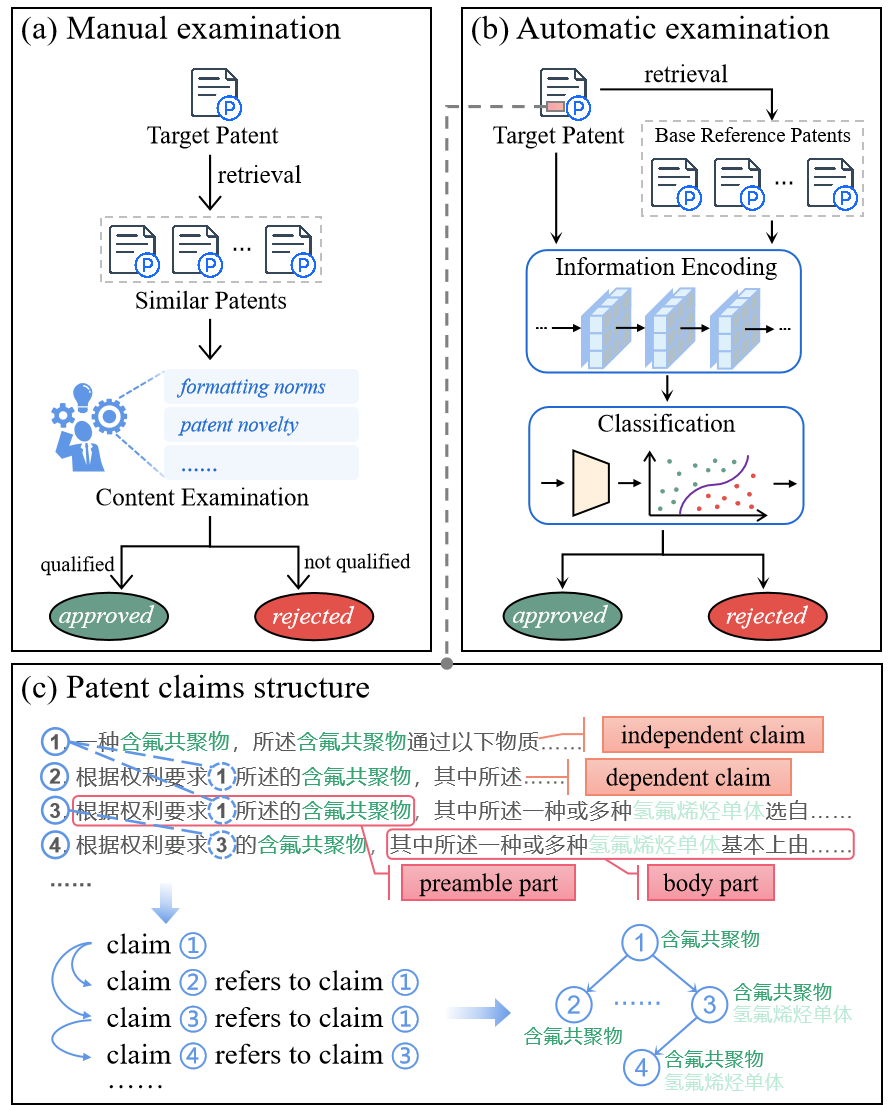}
         \vspace{-6mm}
	\caption{\label{fig_question}
        (a) Manual patent examination for patent approval prediction processes; 
        (b) Automated patent examination based on the manual processes;
        (c) An analysis of the structure within and between patent claims. 
        }
        \vspace{-6mm}
\end{figure}

Rapid advancements in neural networks have enabled neural models to acquire extensive knowledge with growing accuracy and excel in effective similarity retrieval. This progress holds promise for automatic patent examination and approval to overcome the limitations imposed by the examiner's search scope and knowledge reserve. 
However, current research related to patents primarily focuses on single-label or multi-label classification of IPC or CPC numbers \citep{DBLP:journals/corr/abs-1906-02124, DBLP:conf/mlnlp/ChikkamathPOE22}, landscaping classification \citep{DBLP:conf/emnlp/PujariSG0F22}, patent similarity retrieval \citep{DBLP:conf/emnlp/ReimersG19, DBLP:journals/joi/AnLXCS21}, keyword extraction \citep{DBLP:conf/coling/SuzukiT16}, named entity recognition \citep{DBLP:conf/coling/HuV20}, and other related areas. Unfortunately, research is scarce in patent approval prediction, further compounded by the complex nature of the \textit{examination process} and \textit{patent content}. Recalling the complex manual examination process, patent approval involves two primary stages, as illustrated in Figure \ref{fig_question}(a): firstly, \textit{patent retrieval}, where similar patents granted prior to the filing date are obtained as prior art content; secondly, \textit{content examination}, which entails assessing whether the application meets the necessary conditions for approval, including standardized description and innovation compared to the prior art content. This examination process can be approached as a retrieval-based classification task for automated decision-making, as depicted in Figure \ref{fig_question}(b). However, conventional text classification models primarily focus on single-text self-examination \citep{DBLP:conf/aaai/LiZMWYLYS21, DBLP:conf/aaai/LiL0021, DBLP:conf/aaai/GuoH0HL21}, making them unsuitable for direct application in the particular patent approval prediction. 

Additionally, existing patent analysis often dissects complex patent content into the title, abstract, claims, and description, however, resulting in significant information duplication to process efficiently. For example, the title and abstract convey portions of the technical information, claims provide a detailed explanation of the title and abstract, condensing the technical information from the description, while the description is comprehensive but excessively lengthy.
Note that, the importance of claims, which are frequently overlooked, cannot be overstated in patent approval prediction \citep{DBLP:conf/coling/SuzukiT16}. Moreover, claims play a crucial role in defining the protection sought or conferred by a patent application, employing a hierarchical structure. Chinese patent claims, as depicted in Figure \ref{fig_question}(c), distinguishing different key elements by different shades of green, exhibit a rigorous and clear structure. The body part within each claim represents the technical content, while the preamble part establishes the reference relationship and continuously refines the protective scope between different claims. Each claim sentence is represented as a node with a serial number. For instance, nodes 2 and 3 serve as dependent claims, providing a limited description of the key elements in node 1. Also, node 4 refines the key elements of its parent nodes, specifically nodes 1 and 3. 
Intuitively, claims, which encompass rich technical information, concise expression, and notably, structure hierarchy, present an excellent choice for fine-grained patent analysis. However, most of the existing studies have overlooked the importance of claims or often treat them as plain text, disregarding their valuable structural information.

Apart from the potential performance enhancements, incorporating the hierarchical structure of a patent into a retrieval-based classification imitation can facilitate an interpretable process for patent approval. Notable, the approval of patent applications involves a rigorous and transparent decision-making process that necessitates a high degree of evidentiality and traceability. By introducing structural information, patent claims are structurally organized and indexed with the patent representation being enriched with structural prior knowledge, enabling the identification and location of relevant claims through approval backtracking. Precisely, comparing the patent application with existing patents in the library can reveal similarities and dissimilarities in structured claims, reflecting the specification and existing technology of the application, respectively. Only patent applications that are innovative from the prior art and provide correct information are likely to be approved, otherwise not. Therefore, it is necessary and feasible to model the prediction of patent application approval as an evidential decision-making process to ensure transparency, accuracy, and evidentiality throughout the entire process.


\begin{figure*}[htbp]
\vspace{-3mm}
	\centering
        \includegraphics[width=\textwidth]{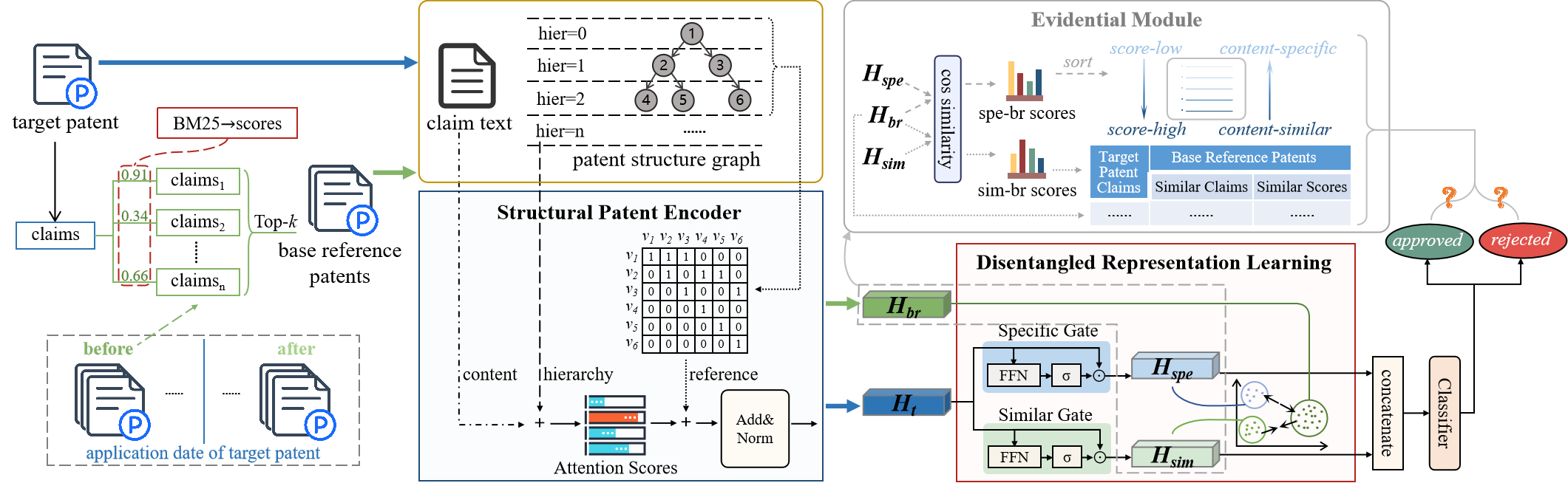}
        \vspace{-7.5mm}
	\caption{\label{fig_model}
        Overall Structure of DiSPat. 
        The blue arrows indicate the transfer of the target patent between modules, and the green arrows indicate the same operation of the base reference patents. }
        \vspace{-4mm}
\end{figure*}

In light of the above discussion, we propose a novel framework DiSPat: structural representation learning and disentanglement for evidential automatic Chinese patent approval prediction. DiSPat includes three primary components: 1) \textit{Base reference retrieval}(BRR) retrieves Top-$k$ most similar patents as a base reference via employing the BM25 algorithm \citep{DBLP:conf/ntcir/MurataKSI05} for fuzzy similarity matching of patent claims. 2) \textit{Structural patent representation}(SPR) learns a structural patent representation to integrate claim text and hierarchy based on a constructed patent structure graph. 3) \textit{Disentangled representation learning}(DRL) disentangles each patent representation into two orthogonal representations that represent the similarity and specificity of patent applications to prior granted patents, respectively. By disentanglement, DiSPat enables interpretable patent comparison and facilitates the identification and location of decision evidence for backtracking.
Finally, a classifier is introduced to predict whether to approve a patent application. In summary, the main contributions of our work are as follows: 

\begin{itemize}[leftmargin=*]
        \vspace{-2mm}
        \item We propose a novel framework of structural representation learning and disentanglement for evidential Chinese patent approval prediction (DiSPat). It marks the first attempt to form automatic patent approval as a retrieval-based classification task and cater to disentangled and structural patent representation for approval accuracy and evidentiality.
        \item We propose SPR to naturally organize the claims of a patent in a graph and subsequently fuse claim text and hierarchy to effectively learn a structural patent representation.
        \item We tailor two contrastive losses to effectively learn disentangled patent representations that incorporate our elaborately designed evidential module to establish a transparent decision-making process of patent approval.
        \item We developed three datasets of Chinese patents from various domains, which are the first of their kind in Chinese and are specified for predicting patent approval. Extensive experiments and case studies on these datasets demonstrate our DiSPat significantly outperforms state-of-the-art baselines on patent approval prediction with improved evidentiality. 
        
\end{itemize}

\vspace{-3mm}
\section{Related Work}

\subsection{Automation Patent Analysis}
In recent years, there has been a growing societal emphasis on intellectual property rights, with a particular surge of interest in patent studies among researchers.
Numerous investigations have delved into various aspects of this domain, such as topic classification employing textual information from the full texts of patents and IPC/CPC labels embedding \citep{DBLP:conf/mlnlp/ChikkamathPOE22, DBLP:conf/birws/PujariMGSF22}, analyzing patents through key sentence extraction and key phrase generation \citep{DBLP:journals/access/SonMLLPJL22}, calculating similarity based on entity and semantic relationship \citep{DBLP:journals/joi/AnLXCS21}, and so on.
These drive further mining of patent information, but little research has been done on patent approvals due to intricate patent comprehension apart from \citep{DBLP:conf/acl/GaoHNZHSK22}. It formulates the novelty scores by comparing each application with millions of prior art using a hybrid of efficient filters and a neural bi-encoder
 to judge the granting of a claim. 
This is an exploration of new fields, but the slightly arbitrary fusion of features makes the decision-making process of patent examination untraceable and non-evidential.

\vspace{-2mm}
\subsection{Patent Representation Learning}
Establishing robust patent representations forms the cornerstone for downstream tasks. 
Presently, widely adopted methods, including LSTM \citep{DBLP:journals/neco/HochreiterS97}, TextCNN \citep{DBLP:conf/emnlp/Kim14}, and Transformer \citep{DBLP:conf/nips/VaswaniSPUJGKP17}, are employed to acquire conventional textual representations.
Within the realm of patents, \citet{DBLP:conf/ecir/PujariFS21} employs SciBERT \citep{DBLP:conf/emnlp/BeltagyLC19}, trained on a scientific publication corpus, to achieve more nuanced patent representations. \citet{DBLP:conf/eeke/ZhangS23} realized the semantic representation of patents by fusing external structural representations such as IPC and plain textual information with GCN. 
However, those studies treat the claims and other text of patents as plain text. 
To harness the inherent structural information within patent texts, 
\citet{DBLP:conf/acl-pitr/FerraroSN14} aims at improving each claim presentation by segmenting it into fragments that then format more clearly, e.g. by adding new lines, 
\citet{DBLP:conf/pacis/LinH10a} builds a tree by parsing claim sentences and extracting the relations between elements, which needs lots of annotation information, 
\citet{DBLP:conf/coling/SuzukiT16} constructs a graph in terms of nodes and connects them by the co-occurrence of terms. 
While these construction methods may be effective for specific tasks, they fall short of obtaining a more rigorous representation based on patent structure, as the overly microscopic hierarchical division may lead to information loss, and more importantly, the inability to index relevant claims by backtracking from a global perspective. 

\vspace{-3mm}
\subsection{Disentangled Representation Learning}
Disentangled representation learning focuses on factorizing the unobservable structural factors from data~\cite{NaseemTZHRN23}. For example, 
creating representations where each dimension is independent and corresponds to a particular attribute has been proposed using VAE variants \citep{DBLP:conf/iclr/HigginsMPBGBML17, DBLP:conf/icml/KimM18}, 
and \citet{DBLP:conf/nips/ChenCDHSSA16} propose a GAN model combined with a mutual information regularization.
Due to its excellent performance and remarkable interpretability, many downstream tasks adopt disentangled representation learning flexibly in various forms, 
\citet{DBLP:conf/acl/BaoZHLMVDC19} generate sentences from disentangled syntactic and semantic spaces, 
\citet{DBLP:conf/sigir/WangJZ0XC20} disentangle the user's intents on adopting the item through the idea of collaborative filtering for more accurate recommendations, 
\citet{DBLP:conf/eccv/SanchezSO20} enforce representation disentanglement based on mutual information to capture the common information between images and specific information of each image. 
This mechanism aligns with our consideration of distinguishing the similarity and specificity of patents, and we believe that it facilitates interpretable patent comparison.

While drawing inspiration from prior works \citep{DBLP:conf/acl/GaoHNZHSK22, DBLP:conf/nips/VaswaniSPUJGKP17, DBLP:conf/acl/BaoZHLMVDC19}, our approach deviates significantly. 
In this paper, we explore the patent approval prediction focusing on claims and probing the inherent structure of these claims. 
To achieve this, we undertake disentanglement to discern key features within the patent, culminating in a predictive classification task.


\begin{table*}
\vspace{-4.5mm}
\caption{\label{Raw Data Statistic}
Raw data information for the three domains A47, C23, and F24.
}
\vspace{-4mm}
\centering
\setlength\tabcolsep{10pt}
\begin{tabular}{cccccc}
\toprule
                                     & \textbf{A47}          &  & \textbf{C23}          &  & \textbf{F24}          \\ \midrule
Max number of claims in a patent     & 50                    &  & 50                    &  & 50                    \\
Average number of claims in a patent & 8                     &  & 9                     &  & 9                     \\
Max length of a claim                & 512                   &  & 512                   &  & 512                   \\
Average length of a claim            & 130                   &  & 112                   &  & 142                   \\ \midrule
Total patents                         & 80778                 &  & 59709                 &  & 86404                 \\ 
Time range                           & 1996/10/28-2023/03/21 &  & 1985/04/01-2023/03/21 &  & 1985/04/01-2023/03/21 \\ \bottomrule
\end{tabular}
\vspace{-2mm}
\end{table*}

\begin{table*}
\caption{\label{Dataset Statistic}
Data statistics for the three datasets A47, C23, and F24.
}
\vspace{-4mm}
\centering
\setlength\tabcolsep{4.8pt}

\begin{tabular}{cccclccclccc}
\toprule
             & \multicolumn{3}{c}{\textbf{A47}} &                      & \multicolumn{3}{c}{\textbf{C23}} &                      & \multicolumn{3}{c}{\textbf{F24}} \\ \cmidrule{2-4} \cmidrule{6-8} \cmidrule{10-12} 
             & Train Set   & Validate Set   & Test Set    &                      & Train Set   & Validate Set   & Test Set    &                      & Train Set   & Validate Set   & Test Set    \\ \midrule
Patent Applications & 46896   & 15632        & 15634   & \multicolumn{1}{c}{} & 33794   & 11264        & 11267   & \multicolumn{1}{c}{} & 50471   & 16823        & 16825   \\
Approved Patents     & 26949   & 8983         & 8984    & \multicolumn{1}{c}{} & 22762   & 7587         & 7588    & \multicolumn{1}{c}{} & 30759   & 10253        & 10253   \\
Approval Rate (\%)  & 57.47   & 57.47        & 57.46   & \multicolumn{1}{c}{} & 67.36   & 67.36        & 67.35   & \multicolumn{1}{c}{} & 60.94   & 60.95        & 60.94   \\ \midrule
Total Patents & \multicolumn{3}{c}{78162}        &                      & \multicolumn{3}{c}{56325}        &                      & \multicolumn{3}{c}{84119}        \\ 
Time Range & \multicolumn{3}{c}{2008/01/01-2023/03/21}        &                      & \multicolumn{3}{c}{2008/01/01-2023/03/21}        &                      & \multicolumn{3}{c}{2008/01/01-2023/03/21}        \\ \bottomrule
\end{tabular}
\vspace{-3mm}
\end{table*}

\vspace{-1mm}
\section{Problem Statement and Datasets}

\subsection{Problem Statement}
Generally speaking, patent approval prediction can be considered a binary classification task. 
For each given patent, we define it as a graph $G=(V, E)$, 
where $V=\{v_{1}, v_2, ..., v_{n}\}$ represents the set of each claim in the patent and $n$ is the number of claims in it, 
while $E=\{e_{ij}|i,j=1, 2,...,n\}$ refers to the set of edges from $v_i$ to $v_j$ which represents the reference relationship between the claims.
Each patent hierarchical graph is associated with a binary label $\hat{Y} \in \{0,1\}$ to denote approval decision, 
where $\hat{Y}=1$ indicates that the patent application is approved and $\hat{Y}=0$ indicates the opposite.

\vspace{-3.5mm}
\subsection{Datasets Preparation}

\textbf{Data Collection.} 
To demonstrate the generalizability of our model to different patent sets in different fields, taking into account sufficient total sample size and a balanced distribution of positive and negative samples, we randomly selected three different categories of fields based on IPC numbers: 
\textbf{A47}, including furniture, domestic articles or appliances, etc.;
\textbf{C23}, including coating metallic material, chemical surface treatment, etc.; 
\textbf{F24}, including heating, ranges, and ventilating.
Through the China National Intellectual Property Administration, we downloaded patents before March 22, 2023, from the results of the three IPC searches respectively. 
Specifically, we obtained approved Chinese patents as positive examples by filtering the item "Patent Type: Invention, Legal Status: Granted" 
and rejected Chinese patents as negative examples by filtering the item "Patent Type: Invention, Legal Status: Rejected". 
The details of raw data are shown in Table \ref{Raw Data Statistic}. 

\noindent\textbf{Data Process.} 
After operations such as similar patent search and patent claims structure analysis, we obtained the processed dataset, see Section \ref{Base Reference Retrieval} and \ref{Hierarchical Patent Structure}. 
The dataset includes the target patents and Top-$k$ base reference patents for each target patent, 
with each base reference patent including claim text and graph structure information, 
and each target patent additionally includes a label.

\noindent\textbf{Data Splits.} 
We randomly divided the dataset into training, validation, and test sets in the ratio of 6:2:2. 
Table \ref{Dataset Statistic} shows the details of the final datasets, 
from which we can see that our three datasets have data sizes of 78162, 56325, and 84119, and their storage sizes are approximately 17.0G, 11.5G, and 20.4G, respectively.

\vspace{-3mm}
\section{Model}
In this section, we introduce our DiSPat for predicting patent approval. 
We first introduced how to retrieve similar patents as base references (Section \ref{Base Reference Retrieval}). 
Then, we describe the method of constructing structural patent representations employing the rigorous structure of patent claims and text information (Section \ref{Structural Patent Representation}). 
We next discuss how to disentangle the obtained structural representations into two orthogonal representations that denote the similarity and specificity of the target patents with the constraints of base reference patents (Section \ref{Disentangled Representation Learning}). 
Finally, we integrate the disentangled outputs to explore patent approval prediction (Section \ref{Patent Approval Prediction}). 

\vspace{-2.5mm}
\subsection{Base Reference Retrieval}
\vspace{-0.5mm}
\label{Base Reference Retrieval}

Patents approved before the filing date express a great deal of prior art, so these patents need to be used as comparative documents in the decision-making process of patent approval. 
In order to ensure that there are enough cases of granted patents as a reference, we judge the approval status of patent applications filed after 2008/01/01 as seen in Table \ref{Dataset Statistic}. 
And for each target patent, we first include all approved patents prior to its application dates in the preliminary selection. 
Then the BM25 algorithm for fuzzy matching based on the claims is used to obtain the Top-$k$ most similar patents in preliminary selection as our base reference patents.

\vspace{-2.5mm}
\subsection{Structural Patent Representation}
\vspace{-0.5mm}
\label{Structural Patent Representation}
In this section, we outline the method for constructing the claims hierarchical graph and obtaining the structural features of the patent.

\vspace{-1mm}
\subsubsection{Hierarchical Patent Structure}\ 
\label{Hierarchical Patent Structure}

The claims of a patent consist of independent and dependent claims, which, in addition to reflecting the technical information of the patent, have an internal hierarchical structure.
Specifically, the independent claims reflect the technical solution of the invention as a whole, and stand on their own; the dependent claims include a preamble part and a body part, in which the preamble part states the number of the reference claim and its subject name, and the body part states the additional technical features of the invention.
The independent claims do not refer to an earlier claim and can be followed by one or more dependent claims, whereas the dependent claims do refer to an earlier claim. 
Generally, the independent claims are written before the dependent claims of the same invention. 

Based on this hierarchy, we extract the hierarchy between the claims to represent the patent in the form of a graph according to the reference part of the dependent claims, using each claim as a node and the reference relationships as edges by the way of Figure \ref{fig_question}(c). 
Since this graph is a tree in our problem, the claims as each node have their hierarchical information. 
As shown in the part of the patent structure graph from Figure \ref{fig_model}, the first independent claim has the lowest hierarchical level, and the other dependent claims have increasing levels of hierarchy based on reference relationships. 
In other words, the independent claim has hierarchy=0, the dependent claims directly referring to it have hierarchy=1, and so on.

\subsubsection{Structural Patent Encoder}\ 

To fuse claim text and patent structure, our model builds upon the Transformer layer \citep{DBLP:conf/nips/VaswaniSPUJGKP17} and incorporates hierarchical embedding and reference embedding. 
In the hierarchy of each patent, each claim is a node, and the original features of a node $v_i$ are obtained by the sum of content embedding and hierarchical embedding:
\begin{equation}
\label{original feature}
\mathbf{h}_i= \operatorname{cont\_emb}(v_i) + \operatorname{hier\_emb}(v_i)
\end{equation}
Content embedding takes advantage of the content of the claim, which we believe contains technical information about the patent. 
We average the BERT token embedding of claim content with dimension $d_h$ as its content embedding. 
Hierarchical embedding is a learnable embedding that takes the hierarchical position of the claim in the patent structure tree as input and outputs a vector with the same dimension as $d_h$. 
To leverage the structural information and the relation between claims, reference embedding modifies the Query-Key product matrix $\mathbf{A}^G$ in the self-attention layer. 
Denote $\mathbf{A}_{i j}^G$ as the $(i,j)$-element of $\mathbf{A}^G$, $\mathbf{V}_{ij}$ as the $(i,j)$-element of value matrix $\mathbf{V}$ projected by value matrices $\mathbf{W}_V^G$, we have: 
\begin{equation}
\label{Aij}
\mathbf{A}_{i j}^G=\frac{\left(\mathbf{h}_i \mathbf{W}_Q^G\right)\left(\mathbf{h}_j \mathbf{W}_K^G\right)^T}{\sqrt{d_h}} + r_{\varphi\left(v_i, v_j\right)}, \quad \mathbf{V}_{ij} = \mathbf{h}_j \mathbf{W}_V^G
\end{equation}
where $\mathbf{h}_i$, $\mathbf{h}_j$ denote the original features of nodes $v_i$, $v_j$ respectively, $ r_{\varphi\left(v_i, v_j\right)} $ denotes the reference embedding, and $\mathbf{W}_V^G \in \mathbb{R}^{d_h \times d_h}$. 
    

The first term in Equation \ref{Aij} is the standard scale dot attention, where query and key are projected by matrices $\mathbf{W}_Q^G \in \mathbb{R}^{d_h \times d_h}$ and $\mathbf{W}_K^G \in \mathbb{R}^{d_h \times d_h}$.
$\varphi\left(v_i, v_j\right): V \times V \rightarrow \mathbb{R}$ is a function which measures the reference relation between $v_i$ and $v_j$ in graph $G$. 
If node $v_j$ refers to node $v_i$, $\varphi\left(v_i, v_j\right)=1$, otherwise 0, 
and $ r_{\varphi\left(v_i, v_j\right)} $ is a learnable scalar indexed by $\varphi\left(v_i, v_j\right)$ and shared across all layers, from which the reference embedding is obtained.  


\begin{table*}[]
\vspace{-4.5mm}
\caption{\label{baselines}
Performance comparison against the baselines on A24, C23, and F24.
}
\vspace{-4mm}
\centering
\setlength\tabcolsep{4.7pt}
\begin{tabular}{cccclccclccc}
\toprule
                                  & \multicolumn{3}{c}{\textbf{A47}} & \multicolumn{1}{c}{} & \multicolumn{3}{c}{\textbf{C23}} & \multicolumn{1}{c}{\textbf{}} & \multicolumn{3}{c}{\textbf{F24}} \\ \cmidrule{2-4} \cmidrule{6-8} \cmidrule{10-12} 
Model                             & ACC (\%)      & AUC (\%)    & Macro F1 (\%)   &                      & ACC (\%)      & AUC (\%)    & Macro F1 (\%)   &                               & ACC (\%)      & AUC (\%)    & Macro F1 (\%)   \\ \midrule
LR(TF-IDF)                                & 65.13    & 61.97    & 61.57      & \multicolumn{1}{c}{} & 67.40    & 52.36    & 50.42      & \multicolumn{1}{c}{}          & 63.42    & 60.28    & 60.42      \\
BiLSTM                            & 65.36    & 62.00    & 55.09      & \multicolumn{1}{c}{} & 70.58    & 62.68    & 63.31      & \multicolumn{1}{c}{}          & 67.57    & 61.64    & 55.71      \\
TextCNN                           & 62.81    & 60.03    & 59.75      & \multicolumn{1}{c}{} & 68.91    & 60.47    & 60.67      & \multicolumn{1}{c}{}          & 60.57    & 56.83    & 56.81      \\
BERT                              & 65.33    & 69.33    & 64.45      &                      & 69.80    & 68.07    & 58.30      &                               & 63.27    & 65.03    & 57.28      \\ \midrule
PatentSBERTa                      & 66.27    & 66.16    & 66.73      &                      & 68.49    & 55.75    & 54.05      &                               & 65.83    & 64.67    & 64.49      \\
TMM                               & 73.86    & 74.18    & 73.65      &                      & 70.84    & 71.80    & 76.27      &                               & 73.40    & 72.97    & 72.52      \\ 
AISeer & 67.97    & 75.09    & 67.50      &                      & 74.16    & 75.75    & 68.62      &                               & 64.38    & 70.59    & 63.32      \\
MMTN & 68.42    & 68.08    & 71.92      &                      & 71.93    & 63.09   & 77.95      &                               & 69.19    & 65.99    & 76.12      \\
HiTIN                             & 74.07    & 74.06    & 75.45      &                      & 80.26    & 73.21    & 75.04      &                               & 71.55    & 73.84    & 71.46      \\ \midrule
DiSPat                              & \textbf{78.34}    & \textbf{85.14}    & \textbf{77.82}      & \multicolumn{1}{c}{} & \textbf{82.16}    & \textbf{88.12}    & \textbf{78.21}      & \multicolumn{1}{c}{}          & \textbf{77.58}    & \textbf{86.19}    & \textbf{77.04}      \\ \bottomrule
\end{tabular}
\vspace{-3mm}
\end{table*}

To calculate the self-attention, $\mathbf{A}^G$ is first processed by applying the Softmax function, which normalizes the values and ensures they sum up to one. Multiplying this normalized attention weight matrix and the value matrix $\mathbf{V}$, then residual connection and layer normalization are applied. 
\begin{equation}
\label{GF_output}
    \mathbf{H} = \operatorname{LayerNorm} (\operatorname{softmax}(\mathbf{A}^G)\mathbf{V}+\mathbf{F})
\end{equation}
where $\mathbf{F} \in \mathbb{R}^{n \times d_h}$ denotes a stacked matrix of all node features. The usual operation is to integrate the features of all the nodes of the input to get a feature of $d_h$ that represents the input patent.
We retain the features of each claim to obtain a representation $\mathbf{H} \in \mathbb{R}^{n \times d_h}$ to denote a patent for subsequent disentanglement and retrospective localization, where $n$ means the number of claims in the patent. 
Thus by the above operation, we obtain the structural representations $\mathbf{H}_t$ and $\mathbf{H}_{br}$, which represent information about the target patent to be applied for and the base reference patents approved already, respectively.


\subsection{Disentangled Representation Learning}
\label{Disentangled Representation Learning}
After obtaining structural patent representations, we turn to disentangle the similarity and specificity of the representation of the target patent. Our inspiration stems from the essence of inventions themselves. A patented invention is a novel technical solution to a product or method. Therefore, the approval of a patent application is not solely determined by its textual content and adherence to format regulations; it is also contingent upon the existing prior art in the respective field.


Therefore, we map the obtained patent representations containing all their claim information into two separate spaces, disentangling them into orthogonal representations indicating the similarity and specificity of the target patents, respectively. 
With the constraint established by the base reference in the preceding Section \ref{Base Reference Retrieval}, we formulate two comparison subtasks for further analysis.

\noindent\textbf{Gate Function.} The gating mechanism serves to regulate the flow and filtering of information, enabling the network to adaptively and selectively pass or filter the input data. The gate function consists of a two-layer FFN: 
\begin{equation}
\mathbf{g}=\sigma(\mathbf{W}_{g2}(\operatorname{ReLU}(\mathbf{W}_{g1} {\mathbf{H}_t}^T + \mathbf{b}_{g1}))+ \mathbf{b}_{g2})
\end{equation}
where $\mathbf{W}_{g1}$, $\mathbf{b}_{g1}$, $\mathbf{W}_{g2}$, $\mathbf{b}_{g2}$ are trainable parameters. The activation $\operatorname{ReLU}$ is applied, and $\sigma$ serves as a non-linear activation function, i.e., sigmoid, for feature screening. $\mathbf{g} \in \mathbb{R}^{n}$ is obtained based on the input representation and through an FFN with sigmoid activation where $n$ is the number of claims in the patent. 
Subsequently, this filtering information is obtained for all claims within a given patent.

\noindent\textbf{Contrastive Losses.} 
Based on the gate mechanism, We design two gates, $\mathbf{g}_{sim}$ and $\mathbf{g}_{spe}$, to get the features that indicate the similarity and specificity of the patent compared to the existing technology by controlling the information transfer, respectively.
\begin{equation}
    \mathbf{H}_{sim} = {\mathbf{g}_{sim}}^T \odot \mathbf{H}_t
\end{equation}
\begin{equation}
    \mathbf{H}_{spe} = {\mathbf{g}_{spe}}^T \odot \mathbf{H}_t
\end{equation}
where $\mathbf{H}_{sim}$ is the similar representation, $\mathbf{H}_{spe}$ is the specific representation and $\odot$ means Hadamard product.
After getting the disentangled representation, we use the base reference set for the constraint to ensure the information accuracy of the disentangled features. 
As we said earlier, the similar feature indicates the part of the target patent that is close to the prior art, so they should be as similar as possible, while the specific feature means the unique innovation of the target patent, so its similarity to the prior art should be as low as possible. 
Therefore, two contrastive losses are tailored to constrain the similarities:
\begin{equation}\label{L_sim}
    \mathcal{L}_{con}^{sim} = \sum  F_{SIM}( \mathbf{H}_{sim}, \mathbf{H}_{br}) 
\end{equation}
\begin{equation}\label{L_spe}
    \mathcal{L}_{con}^{spe} = \sum  F_{SIM}( \mathbf{H}_{spe}, \mathbf{H}_{br}) 
\end{equation}
\begin{equation}\label{L_con}
    \mathcal{L}_{con} = \mathcal{L}_{con}^{sim} + \mathcal{L}_{con}^{spe}
\end{equation}
where $\mathbf{H}_{br}$ denotes the features of all patents in the base reference set of this target patent. $F_{SIM}$ is the similarity calculation function pushing the disentangled representation to converge to orthogonality, which is defined as follows:
\begin{equation}
    F_{SIM}(x_1, x_2)=\left\{\begin{array}{c}
    1 - \cos (x_1, x_2), \operatorname{if} \, x_1=\mathbf{H}_{sim} \\
    \cos (x_1, x_2), \,\,\, \operatorname{if} \, x_1=\mathbf{H}_{spe}
\end{array}\right.
\end{equation}
where $\operatorname{cos}$ denotes the cosine similarity calculation, 
$x_1$ and $x_2$ denote the two representations input to the function $F_{SIM}$. 
Equation \ref{L_sim} and Equation \ref{L_spe} enter $\mathbf{H}_{sim}$ and $\mathbf{H}_{spe}$ as $x_1$, respectively, and both enter $\mathbf{H}_{br}$ as $x_2$ when calling $F_{SIM}$.



\subsection{Patent Approval Prediction}
\label{Patent Approval Prediction}
This paper treats the approval prediction task as a binary classification problem. Accordingly, we integrate similar and specific features of the target patent for prediction and apply the full connection layer and activation layer to calculate the label $\hat{y}$: 
\begin{equation}
\hat{y} = \operatorname{softmax} ( \mathbf{W}_p [\mathbf{H}_{sim}, \mathbf{H}_{spe}]  + \mathbf{b}_p)
\end{equation}
where $\mathbf{W}_p$ and $\mathbf{b}_p$ are trainable parameters, and $[\cdot,\cdot]$ denotes concatenation operation. 

In the next step, for the given patent application, the model is trained to minimize the cross-entropy value between the predicted probability and the ground truth:
\begin{equation}
\label{L_clf}
     \mathcal{L}_{clf} = -\mathbb{E}_{y \sim \hat{Y}}[y \text{log}(\hat{y}) + (1-y)\text{log}(1- \hat{y})]
\end{equation}
where $y$ and $\hat{y}$ represent the ground-truth and the predicted probability respectively, while $\hat{Y} \in \{\text{1: Approved, 0: Rejected}\}$ is a set of ground truth labels.  

The overall training loss is a combination of the contrastive losses (\ref{L_sim}-\ref{L_con}) and the classification loss (\ref{L_clf}), i.e., minimizing 
\begin{equation}
\label{L_all}
    \mathcal{L}  = \mathcal{L}_{con} + \mathcal{L}_{clf} = \mathcal{L}_{con}^{sim} + \mathcal{L}_{con}^{spe} + \mathcal{L}_{clf} 
\end{equation}

\section{Experiments}
In this section, we evaluate the effectiveness of our proposed model on our machine and three datasets.

\vspace{-2mm}
\subsection{Baselines and Settings}

\subsubsection{Baselines}\

Because the patent approval prediction problem has rarely been considered in previous studies, several common text classifiers and state-of-the-art classification models in different domains are chosen as baselines to compare with our proposed method. 
The common text classifiers are as follows: 
\begin{itemize}[leftmargin=*]
	\item \textbf{LR(TF-IDF)} refers to logistics regression using TF-IDF features with Adam optimizer (2e-5), 10 epochs' run, and batch size of 16.

        \item \textbf{BiLSTM}
        \citep{DBLP:journals/neco/HochreiterS97} with Glove \citep{DBLP:conf/emnlp/PenningtonSM14} embeddings as the input. 
        The batch size is set to 32. 
    
	\item \textbf{TextCNN}
        \citep{DBLP:conf/emnlp/Kim14} with Glove \citep{DBLP:conf/emnlp/PenningtonSM14} embeddings as the input.
        The batch size is set to 50. 
    
	\item \textbf{BERT}
        \citep{DBLP:conf/naacl/DevlinCLT19} with fine-tuning in patent classification. 
        The batch size is set to 256. 
    
\end{itemize}

In addition, several advanced classification models in patent and other domains are chosen as baselines:
\begin{itemize}[leftmargin=*]
        \item \textbf{PatentSBERTa}(\citeyear{bekamiri2021patentsberta}) \citep{bekamiri2021patentsberta}, using patent claims text to calculate patent-to-patent (p2p) similarity and thus CPC classification. 
        The batch size is set to 512. 
        
        \item \textbf{TMM}(\citeyear{DBLP:conf/ecir/PujariFS21}) \citep{DBLP:conf/ecir/PujariFS21}, 
        concatenating the patent's title and abstract as input and using SciBERT model (trained on a corpus of scientific publications) \citep{DBLP:conf/emnlp/BeltagyLC19} for IPC/CPC classification. 
        The batch size is set to 2.
        
        \item \textbf{AISeer}(\citeyear{DBLP:conf/acl/GaoHNZHSK22}) \citep{DBLP:conf/acl/GaoHNZHSK22}, the first model aiming to predict the approval task of a single claim. 
        The batch size is set to 128. 

        \item \textbf{MMTN}(\citeyear{wang2023positive}) \citep{wang2023positive}, a fake news detection model that uses a masked Transformer to strengthen the correlations between relevant content. 
        The batch size is set to 16. 

        \item \textbf{HiTIN}(\citeyear{DBLP:conf/acl/ZhuZHW023}) \citep{DBLP:conf/acl/ZhuZHW023}, enhancing the text representation for hierarchical text classification with structural information of the label hierarchy.
        Because there is no hierarchical relationship between patent approval results, we set the labels' parent information to null and the batch size is set to 8. 
        
\end{itemize}

\begin{table*}[!t]
\vspace{-4mm}
\caption{\label{tab_ablation_SPR}
Performance with variants of SPR on datasets of A47, C23, and F24. 
\textit{w/o} stands for \textit{without} and \textit{r.p.} stands for \textit{replace}. 
}
\vspace{-4mm}
\centering
\setlength\tabcolsep{4.1pt}
\begin{tabular}{lccccccccccc}
\toprule
                    & \multicolumn{3}{c}{A47}                          &  & \multicolumn{3}{c}{C23}                          &  & \multicolumn{3}{c}{F24}                          \\ \cmidrule{2-4} \cmidrule{6-8} \cmidrule{10-12} 
Variants of SPR     & ACC (\%)          & AUC (\%)          & Macro\ F1 (\%)    &  & ACC (\%)          & AUC (\%)          & Macro\ F1 (\%)    &  & ACC (\%)          & AUC (\%)          & Macro\ F1 (\%)    \\ \midrule
Base architecture   & \textbf{78.34} & \textbf{85.14} & \textbf{77.82} &  & \textbf{82.16} & \textbf{88.12} & \textbf{78.21} &  & \textbf{77.58} & \textbf{86.19} & \textbf{77.04} \\
\textit{-w/o} hier embedding & 74.35          & 84.02          & 74.32          &  & 81.58          & 87.31          & 77.24          &  & 72.94          & 85.93          & 72.87          \\
\textit{-w/o} ref embedding  & 76.17          & 84.83          & 76.07          &  & 81.64          & 88.06          & 76.97          &  & 72.45          & 86.39          & 72.41          \\
\textit{-r.p.} fc graph      & 74.71          & 83.97          & 74.65          &  & 81.81          & 87.97          & 76.98          &  & 76.07          & 85.98          & 75.74          \\ \bottomrule
\end{tabular}
\vspace{-3mm}
\end{table*}

\begin{figure*}[!t]
	\centering
        \includegraphics[width=0.95\linewidth,trim=0 4 0 30,clip]{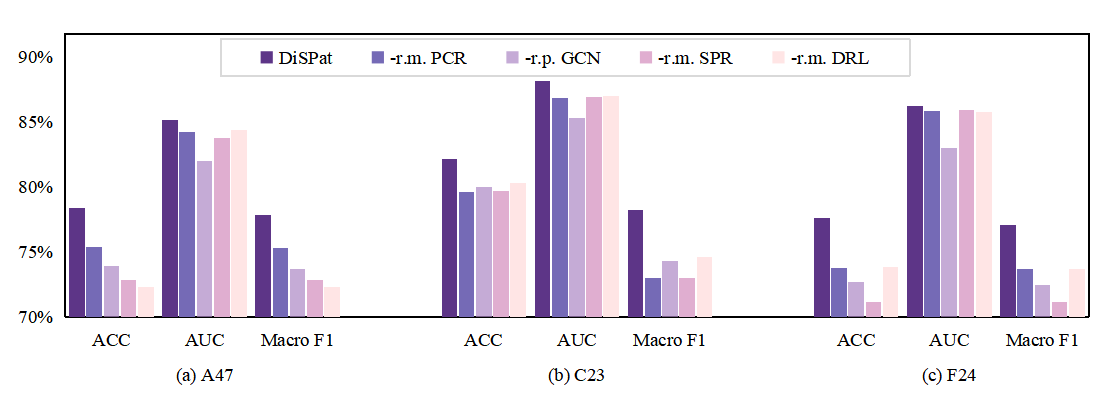} 
        \vspace{-5mm}
	\caption{\label{fig_ablation_global}
         Performance when replacing or removing certain components from DiSPat on the three datasets. 
         \textit{r.p.} stands for \textit{replace} and \textit{r.m.} stands for \textit{remove}. }
         \vspace{-4mm}
\end{figure*}

For a fair comparison, we run all models on our datasets and machine. 
Since our prediction task focuses on the claims of patents, we replace the inputs of all models with claims, and these models are kept with their claim-only version. 
Also to fit our dataset and machine runnability, we adjust the batch size of the baseline models accordingly, 
but keep the rest of the optimizers, learning rates, etc. the same as the original baseline models.

\vspace{-1mm}
\subsubsection{Settings}\ 

In the data preparation section of our experiment, for selecting hyper-parameters the Top-$k$ most similar patents as a base reference, the max number of claims $n$ in a patent, and the max number of words $w$ in a claim, we consider values from $\{1,3,5,7,9\}$, $\{5,10,15,20,30,40,50\}$ and $\{32,64,128,256,512\}$ respectively. 
Ultimately, we set $k$, $n$, and $w$ as 5, 20, and 512 respectively.
We get the processed dataset by the BM25 algorithm and then use PyTorch to implement the models. 
Bert obtains content embedding with Transformers \citep{DBLP:conf/emnlp/WolfDSCDMCRLFDS20} as the base architecture. 
For the structural patent encoder, we set the number of attention heads and feature sizes $d_h$ as 6 and 768 respectively. 
The FFNs in DRL are all multi-layer FC networks with the number of layers set to 2. 
The batch size is set to 4. 
The optimizer utilizes Adam with a learning rate of 1e-4. 
The dropout rate is set to 0.1. 
We train our model on a single NVIDIA GeForce RTX 4090 GPU for 50000 steps and save the model every 5000 steps, through which we get the best-performance model based on evaluation metrics ACC (Accuracy)/AUC (Area under the ROC Curve)/Macro F1 on the train set and reload it for the evaluation on the test set. 
We report the average testing results over five runs.  




\begin{figure*}[!t]
\vspace{-6mm}
	\centering
        \subfigure[Different Top-$k$ most similar patents]
        {\label{fig_parameter_acc_k}
            \includegraphics[width=0.3\linewidth,trim=0 11 0 0,clip]{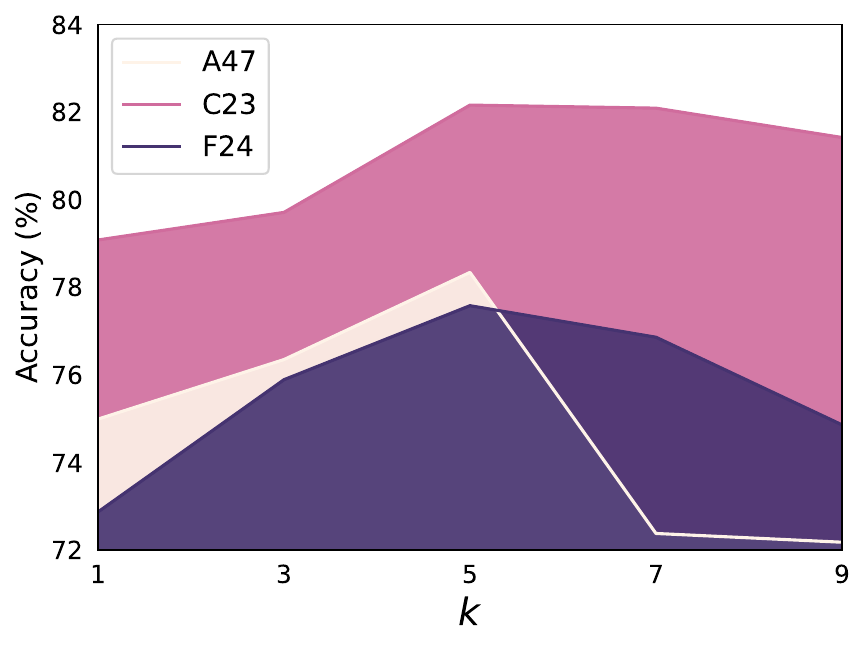}
        }
	\subfigure[Different number of claims $n$ in a patent]         
        {\label{fig_parameter_acc_n}
            \includegraphics[width=0.3\linewidth,trim=0 11 0 0,clip]{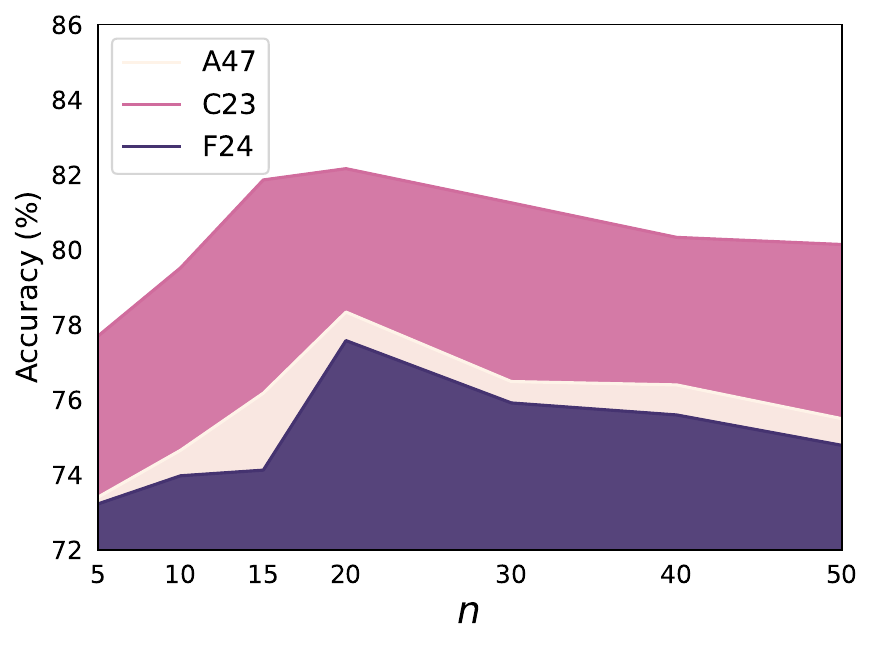}
        }
        \subfigure[Difference number of words $w$ in a claim]{\label{fig_parameter_acc_w}
            \includegraphics[width=0.3\linewidth,trim=0 11 0 0,clip]{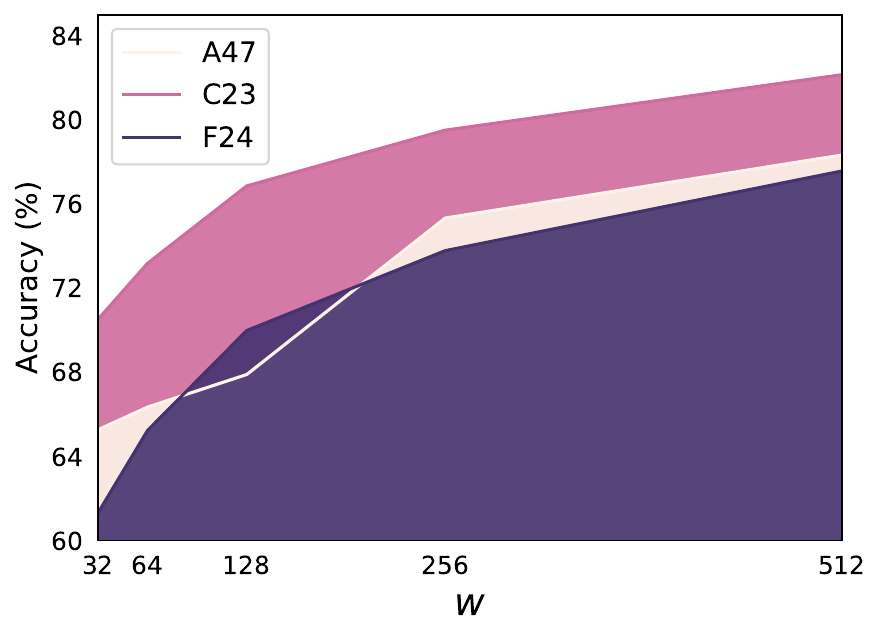}
        }
    \vspace{-5mm}
	\caption{\label{fig_parameter}
        The results of comparison with different values of $k$, $n$, and $w$ on datasets A47, C23 and F24.}
    \vspace{-4mm}
\end{figure*}

 \vspace{-4mm}
\subsection{Overall Performance}

The performances of the different models in terms of ACC, AUC, and Macro F1 are listed in Table \ref{baselines}. 
The results indicate that our DiSPat outperforms other baselines, which proves the effectiveness of the patent hierarchical structure and disentanglement mechanism. 

From the results, we can make the following observations:

(1) Although LR(TF-IDF), BiLSTM, TextCNN, and BERT are efficient, they have significant limitations in terms of the 
prediction accuracy since they only use the textual features of claims.

(2) PatentSBERTa and AISeer introduce similarity measures based on p2p and patent novelty, respectively, their improved performance over traditional models illustrates the importance of base reference retrieval in patent approval prediction tasks, providing a basis for analyzing the similarity and specificity of target patents in our DiSPat; 
MMTN uses a masked Transformer to filter the noise between claims, and it is evident that the structural prior knowledge from patent claims has a great role in potential performance enhancements.
TMM uses pre-trained models from the field of scientific literature, HiTIN fuses text and label features, and they make performance improvements by leveraging external information other than claim text.

(3) Dataset C23 performs better compared to the other two datasets, which may be due to the fact that dataset C23 in the chemical domain has the shortest average length of claims (as can be seen through Table \ref{Raw Data Statistic}. The relatively condensed expression and repetition of some compounds in the patents in the chemical domain reduces the impact of noise and outliers on the training and helps the model comprehend the text content more accurately.

(4) We can observe that our proposed DiSPat achieves state-of-the-art performance on all three datasets. 
In a departure from previous studies, DiSPat integrates text and hierarchy based on a constructed patent structure graph to explore the enrichment of patent representations.
In addition, according to the property of invention patents, the DiSPat disentangles each patent representation into two orthogonal representations that represent the similarity and specificity of patent applications to prior granted patents, respectively, enabling it to provide necessary and feasible information for downstream prediction tasks.

\vspace{-1mm}
\subsection{Ablation Study}

To analyze the contribution of each component of DiSPat, we remove or replace these components to test our model, and the results are in Figure \ref{fig_ablation_global}. 
We first remove BRR (\textit{-r.m.} BRR), and it can be seen that the disentangled representation lacks the corresponding constraints without the base reference, and this single-text self-examination process would be much less effective in this task. 
Whether replacing our structural patent encoder with GCN (\textit{-r.p.} GCN) or removing it (\textit{-r.m.} SPR) entirely leaving only BERT to obtain textual representations, we find that our encoder outperforms them on our task, and this stem from the lack of rigorous and clear claim construction information. 
Removing DRL (\textit{-r.m.} DRL) represents that after getting the structural patent representation, we transform the model into one that does not take into account similarity to existing patents and its specification, but rather direct classification. 
Its effect declined somewhat compared to DiSPat, illustrating the importance of retrieval-based settings and tailored contrastive losses for the task of patent prediction. 

We further analyze the effect of structural patent representation and disentangled representation learning in detail.

\subsubsection{Analysis of Structural Patent Representation}\ 
\vspace{1mm}

Regarding the components of SPR, we validate the utility of hierarchical embedding, reference embedding, and the structural construction information of the patent. 
The empirical results of the model variants on different datasets are summarized in Table \ref{tab_ablation_SPR}. 
By removing hierarchical embedding (\textit{-w/o} hier embedding) and reference embedding (\textit{-w/o} ref embedding), the performance degradation demonstrates the effectiveness of structural prior knowledge, such as the hierarchical position of the claims and the references between the claims, in enriching patent representation. 
When replacing the claims hierarchy graph with a fully connected graph (\textit{-r.p.} fc graph) before disentangling and classifying them, the performance decreases. 
The results indicate the structural claim graph is valuable for obtaining patent features that precisely express technical and protection information for fine-grained patent analysis.

\subsubsection{Analysis of Disentangled Representation Learning}\ 

\begin{figure}[h]
\vspace{-2mm}
	\centering
	\includegraphics[width=0.8\linewidth]{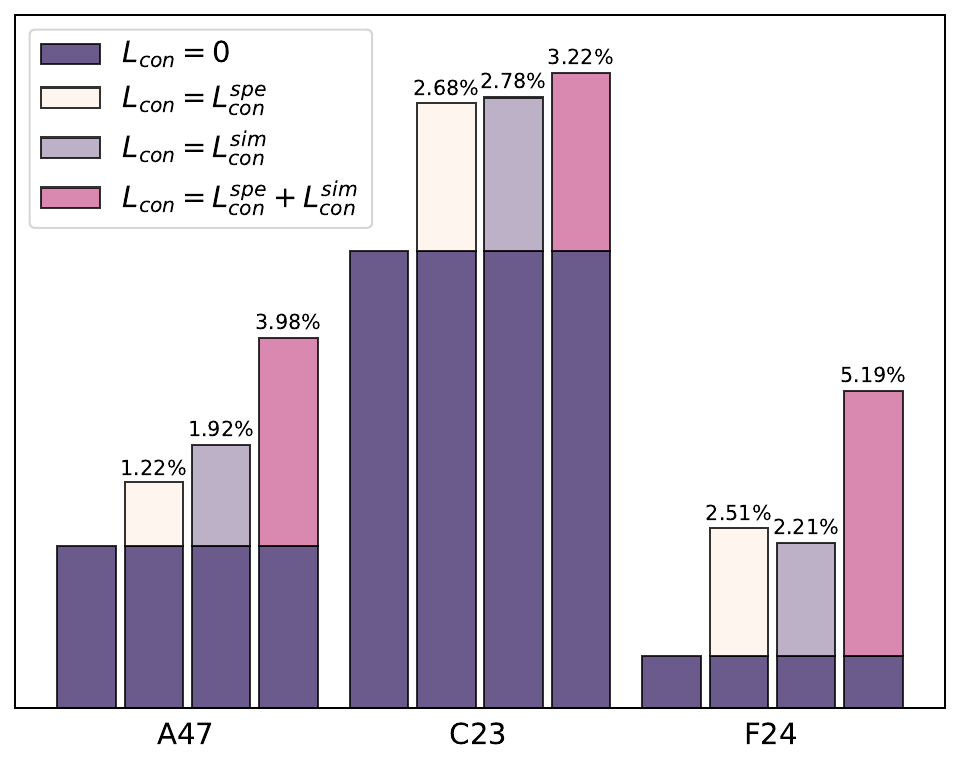}
    \vspace{-5mm}
	\caption{\label{fig_ablation_DRL_acc}
        Impact of different $\mathcal{L}_{con}$ compositions on accuracy in DRL module. 
        The percentage indicates the improvement of the accuracy compared to $\mathcal{L}_{con}=0$. 
        }
    \vspace{-2mm}
\end{figure}

The impacts of different $\mathcal{L}_{con}$ compositions in DRL on three datasets are shown in Figure \ref{fig_ablation_DRL_acc}, and we take $\mathcal{L}_{con}=0$ as the underline to see the performance improvement magnitude of the other approaches compared to it. 
$\mathcal{L}_{con}=0$ is the approach that disregards similarity and specificity, in other words, ignoring the base reference; 
$\mathcal{L}_{con}=\mathcal{L}_{con}^{spe}$ only uses the base reference to constrain specific representations; 
$\mathcal{L}_{con}=\mathcal{L}_{con}^{sim}$ only uses the base reference to constrain similar representations; 
and $\mathcal{L}_{con}=\mathcal{L}_{con}^{spe}+\mathcal{L}_{con}^{sim}$ is our base architecture which is DiSPat.

On all datasets, we can observe through $\mathcal{L}_{con}=0$ that the base reference is essential for disentangling constraints, which in turn affects the prediction results.
And the performance of $\mathcal{L}_{con}=\mathcal{L}_{con}^{spe}$ is similar to that of $\mathcal{L}_{con}=\mathcal{L}_{con}^{sim}$, better than $\mathcal{L}_{con}=0$, but both are not as outstanding as DiSPat, which demonstrates the indispensability of tailored contrastive losses and considering only similarity or specificity is one-sided to predict patents' approvement.

\begin{figure*}[!t]
\vspace{-7mm}
	\centering
        \subfigure[an approved patent application]{\label{fig_explain_1_a}
            \includegraphics[width=0.45\linewidth,trim=0 2 0 0,clip]{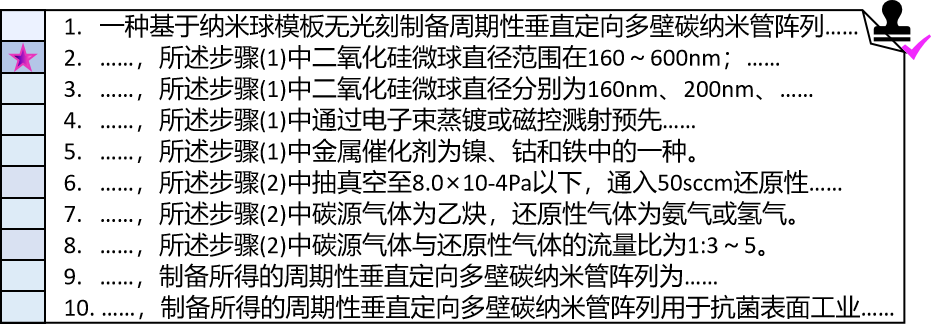}
        }
	\subfigure[a rejected patent application]             {\label{fig_explain_1_b}
            \includegraphics[width=0.45\linewidth,trim=0 2 0 0,clip]{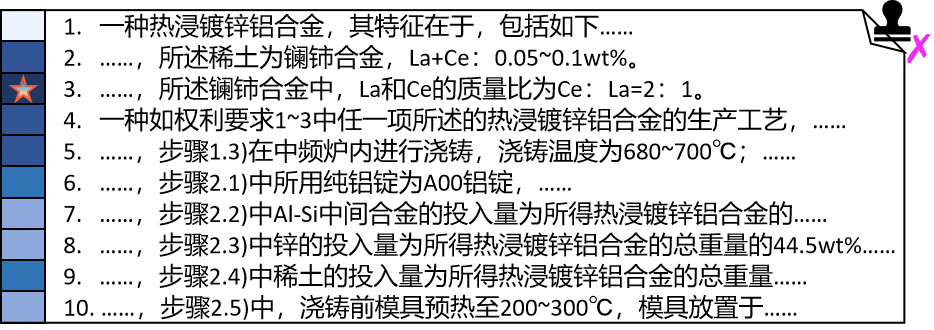}
        }
    \vspace{-6mm}
	\caption{\label{fig_explain_1}
        Claim specification visualization of positive and negative instances. 
        Lighter-colored cells indicate lower $\mathbf{H}_{spe}$ to $\mathbf{H}_{br}$ similarity scores, indicating stronger specificity. 
        The stars denote the claims with the highest similarity score.
        }
    \vspace{-6mm}
\end{figure*}

\begin{figure*}[!t]
	\centering
        \subfigure[an approved patent application]{\label{fig_explain_2_a}
            \includegraphics[width=0.45\linewidth,trim=0 2 0 0,clip]{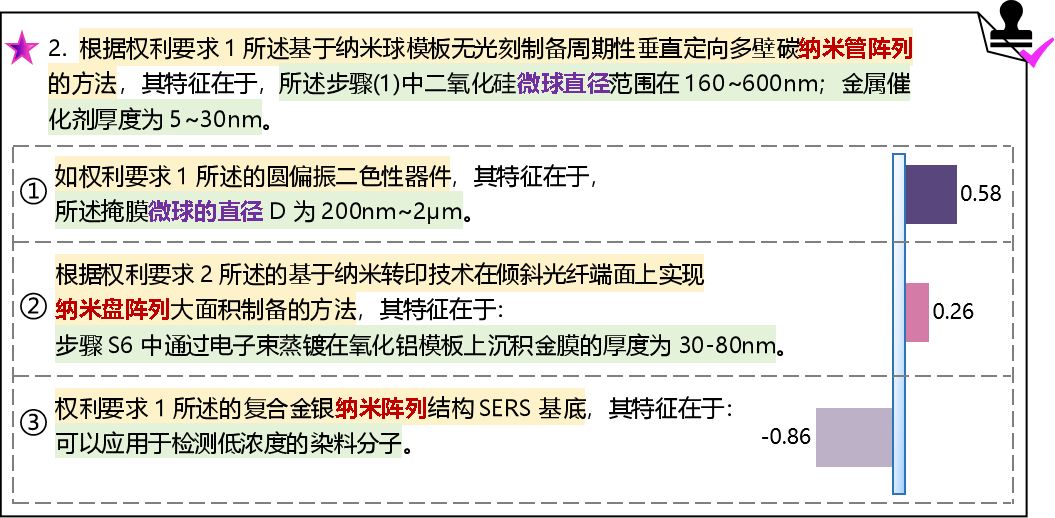}
        }
	\subfigure[a rejected patent application]{\label{fig_explain_2_b}
            \includegraphics[width=0.45\linewidth,trim=0 2 0 0,clip]{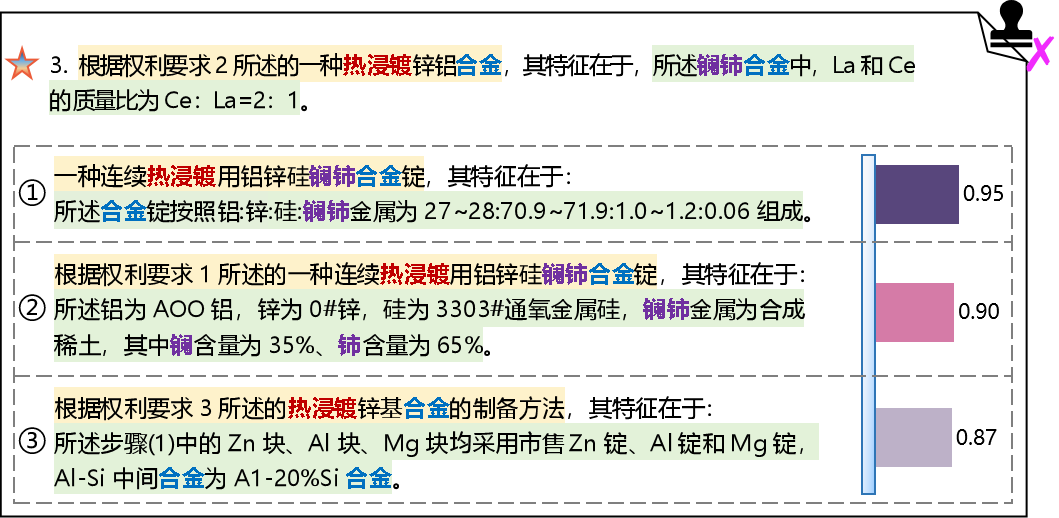}
        }
    \vspace{-5.5mm}
	\caption{\label{fig_explain_2}
        Examples of patent application backtracking. 
        The yellow area and the green area denote the preamble and body parts of the claim, respectively. 
        Inside the dashed box are 3 similar claims found for the above content (marked by stars). 
        Different key phrases are represented by different colored bolded faces, and the bars denote the similarity scores of $\mathbf{H}_{sim}$ and $\mathbf{H}_{br}$.
        }
    \vspace{-4mm}
\end{figure*}

\subsection{Parameter Sensitivity Analysis} 

Different Top-$k$ most similar base references to the target patent, a crucial element determining the scope of reference comparison, hold significance in patent approval prediction. 
The experiment results shown in Figure \ref{fig_parameter_acc_k} indicate that, as the increase of $k$, the performance of all datasets first increases and then decreases, which means that more base references provide more valid information until a certain threshold is reached, after which irrelevant noise is introduced due to the low similarity between the base reference patents and the target patent, degrading performance.
The best numbers of base references on three datasets are all 5.

In Figure \ref{fig_parameter_acc_n}, we discuss the effect of the number of claims $n$ in a patent on the prediction, and it is seen that the best results are all obtained when $n=20$.
Statistically, the distribution of data in our three datasets is essentially the same. The max values of $n$ are 50, with an average of 8 to 9 claims per patent  (as shown in Table \ref{Raw Data Statistic}), and the percentage of patents with $n<=20$ is over 95\%. 
Therefore, we think that incomplete technical information biases the prediction when $n<20$. 
Conversely, an escalating percentage of padding diminishes information density, mitigating model performance when $n>20$, and concurrently extending runtimes. 

To investigate the influence of the max number of words in a claim $w$, we set several experiments, and the performance exhibited in Figure \ref{fig_parameter_acc_w} shows that the accuracy increases with $w$ on our datasets. 
According to statistics, the percentages of $w<=128$ in A47, C23, and F24 datasets are 70.39\%, 76.31\%, and 64.27\%, respectively, and the claims with other lengths are more evenly distributed, with the longest being 512 words. 
This suggests that all parts of the claims, including the preamble and body parts, are necessary and provide comprehensive technical information, which also confirms that our choice of claims for patent approval prediction is sound. 

\subsection{Evidential Analysis}

To establish the transparent decision-making process of patent approval, we conducted the following experiment elaborately, as shown in the Evidential Module of Figure \ref{fig_model}.



To intuitively demonstrate the difference in characteristics of target patent captured by disentangled representation learning (DRL) in approved and rejected patents, we visualize the similarity scores of $\mathbf{H}_{spe}$ with $\mathbf{H}_{br}$, with the lighter-color cells showing lower similarity scores and proving stronger specificity. 
A contrastive case study is presented in Figure \ref{fig_explain_1}. 
In this case, there are more lighter-colored claims in an approved application than in a rejected one as expected, which means that our model does learn specific information on the one hand, 
and there is more specific and less similar technical information contained in the approved application compared with the base reference on the other hand. 
The parts of a patent application that are more similar to the past may often play a very important role in determining whether it is approved or rejected, 
therefore we denote the two claims with the highest similarity scores in the two cases of Figure \ref{fig_explain_1} by stars to focus on.

Disentangled features are important, especially for rejected patent applications, 
cooperating with indexable structural representation to identify and locate the relevant claims through approval backtracking. 
By calculating the similarity between the $\mathbf{H}_{sim}$ and $\mathbf{H}_{br}$, we get the similarity scores and from where we go to find the claims that are most similar to each claim sentence in the target patent. 
We perform this operation for the two claims starred, those are, Claim 2 in Figure \ref{fig_explain_1_a} and Claim 3 in Figure \ref{fig_explain_1_b}.
These two claims are also shown in the upper portion of Figure \ref{fig_explain_2}, and we can see some of the existing claims that are similar to them from the base reference set and the similarity scores in the dotted boxes.
As can be seen from this case of an approved patent in Figure \ref{fig_explain_2_a}, the similarity scores are generally small and the key phrases are unable to hit both the preamble and the body of a sentence of claim, which suggests that this patent is developing new approaches for the subjecting matter or migrating methods those have already been used in other fields, and therefore do not affect the novelty of the patent.
Unlike it, the case of a rejected patent (as seen in Figure \ref{fig_explain_2_b}) where the key phrases hit both the preamble and body parts, in addition, the similarity scores are generally high, suggesting that the patent is conflicting with the prior art in its field, which solving the existing problem of figuring out why the patent is rejected.

By mining and reasoning these clues, the model ensures transparency, accuracy, and evidentiality throughout the entire decision-making process of patent approval. 

\vspace{-3mm}
\section{Conclusions}
In this paper, we propose a novel framework DiSPat: structural representation learning and disentanglement for evidential Chinese patent approval prediction. 
We model this task as an evidential decision-making process, and the base reference retrieval module (BRR) is developed to provide similar prior granted patents. 
To effectively learn a structural patent representation, we introduce SPR to naturally organize the claims of a patent in a graph and subsequently fuse claim text and hierarchy. 
A transparent examination process of patent approval is established through two tailored contrastive losses in DRL to learn orthogonal disentangled patent representations considering similarity and specificity. This approach enables interpretable patent comparison and facilitates the identification and location of decision evidence for backtracking. 
Extensive experiments and case studies on our original datasets demonstrate that our DiSPat significantly outperforms state-of-the-art baselines in terms of both performance and evidentiality. 

\vspace{-3mm}
\section{Acknowledgements}
This work is supported by National Natural Science Foundation of China (No. 62372043). 

\bibliographystyle{ACM-Reference-Format}
\balance 
\bibliography{sample-base}


\end{document}